\documentclass[twocolumn, 12pt]{webofc}
\usepackage[varg]{txfonts}  
\usepackage[colorlinks]{hyperref}
\usepackage{graphicx}
\usepackage{caption}
\usepackage[skip=0.5ex]{subcaption}

\makeatletter
\renewcommand{\fnum@figure}{Fig. \thefigure}
{

\begin{document}

\title{ \LARGE Giant amplification of Berezinskii-Kosterlitz-Thouless transition temperature in superconducting systems characterized by cooperative interplay of small-gapped valence and conduction bands }

\author{\firstname{Giovanni} \lastname{Midei}\inst{1}\inst{2}\fnsep\thanks{\email{giovanni.midei@unicam.it}} \and
        \firstname{Andrea} \lastname{Perali}\inst{3}\fnsep\thanks{\email{andrea.perali@unicam.it}}        
}

\institute{School of Science and Technology, Physics Division, University of Camerino,
Via Madonna delle Carceri, 9B, Camerino (MC), Italy
\and
INFN-Sezione di Perugia, 06123 Perugia, Italy
\and
 School of Pharmacy, Physics Unit, University of Camerino,
Via Madonna delle Carceri, 9B, Camerino (MC), Italy
}
\abstract{\\
Two-dimensional superconductors and electron-hole superfluids in van der Waals heterostructures having tunable valence and conduction bands in the electronic spectrum are emerging as rich platforms to investigate novel quantum phases and topological phase transitions. In this work, by adopting a mean-field approach considering multiple-channel pairings and the Kosterlitz-Nelson criterion, we demonstrate giant amplifications of the Berezinskii-Kosterlitz-Thouless (BKT) transition temperature and a shrinking of the pseudogap for small energy separations between the conduction and valence bands and small density of carriers in the conduction band. The presence of the holes in the valence band, generated by intra-band and pair-exchange couplings, contributes constructively to the phase stiffness of the total system, adding up to the phase stiffness of the conduction band electrons that is boosted as well, due to the presence of the valence band electrons. This strong cooperative effect avoids the suppression of the BKT transition temperature for low density of carriers, that occurs in single-band superconductors where only the conduction band is present.  Thus, we predict that in this regime, multi-band superconducting and superfluid systems with valence and conduction bands can exhibit much larger BKT critical temperatures with respect to single-band and single-condensate systems. 
}

\maketitle

\section{Introduction}
\label{intro}
Two dimensional multi-band superconductivity can generate rather interesting physics \cite{Milosevic2015}, especially in the case of electronic systems with valence and conduction bands both participating to the superconducting condensate. In such configurations, complex redistribution of electrons occurs between valence and conduction bands, leading to a new physics with respect to single-band superconductors, such as density induced and band-selective BCS-BEC crossover \cite{Andrenacci1999}, topological quantum phase transitions and hidden criticalities \cite{Nozieres1999,sademelo, Yue2022, Litak2012}. Furthermore, at finite temperature, this phenomenon is also responsible for the non-monotonic behavior of the superconducting gaps resulting in a superconducting-normal state reentrant transition \cite{Midei2023}.
A peculiar feature about two dimensional systems regards the nature of the superconducting transition. Indeed, in thin films the transition to the superconducting state with decreasing temperature occurs in two stages: at first, the finite amplitude of the order parameter forms around the mean-field critical temperature $T_{MF}$ but without ordering of the phase, after that, the true superconducting transition occurs with phase ordering at a lower temperature, which is expected to belong to the universality class of the BKT transition of the two dimensional XY model \cite{Berezinskii1972, Kosterlitz1973, Kosterlitz1974}. In contrast to the power-law dependence of the coherence length $\xi_c$ predicted in the Ginzburg-Landau theory \cite{Benfatto2000, Mondal2011}, this transition is characterized by an exponential divergence at the BKT critical temperature. In order to observe the BKT phenomenon, a condition is that $d << \xi_{c}$ being $d$ the sample thickness, though some experiments have reported the BKT transition also outside the expected ranges \cite{Chu2004}. 
In this regime, the transition to the normal state is driven by the vortex-antivortex dissociation instability, which is connected with the logarithmic dependence of the interaction energy on the separation between vortices. This leads to the discontinuous jump of the phase stiffness $J$ from a finite value right below $T_{BKT}$ to zero above it \cite{Nelson1977, Minnhagen1987}. The value of $J$ at the BKT critical temperature can be inferred via measurements of the London penetration depth or from the nonlinear exponent of the I-V characteristics \cite{Venditti2019, Sharma2022}. 
The study of the BKT transition in valence and conduction band systems can be relevant for recently discovered 2D superconducting bilayer graphene systems \cite{Zhou2022, Pantaleon2022}, where the energy shift between the conduction and valence bands can be 
precisely tuned by an external electric field perpendicular to the graphene layers, and electron-hole superfluid systems, such as the double-bilayer graphene (2BLG) \cite{Perali2013}, consisting of two conducting bilayer graphene sheets, one containing electrons and the other holes. The two bilayer sheets are separated by an insulating barrier in order to prevent tunneling between the sheets. This system can be studied by means of a BCS mean-field theory by performing a standard particle-hole transformation \cite{Conti2017}. 
In this way, the intra-band couplings create Cooper pairs made up by a positively charged hole in the conduction/valence band of the p-doped bilayer and an electron in the conduction/valence band of the n-doped bilayer, while the pair-exchange couplings transfer electrons and holes involved in the formation of Cooper pairs from the conduction band to the valence band (and viceversa) of their respective graphene bilayer. Furthermore, materials where charge density wave (CDW) and superconducting orders coexist can be an interesting platform for studying valence and conduction band configuration, such as underdoped cuprates \cite{Gabovich2009, Gabovich2010, Arpaia2019, Perali1996} and transition metal dichalcogenides \cite{Rossnagel2011,Neto2001,Kiss2007}. In fact, the presence of CDWs and their fluctuations can modify the energy spectrum by opening (pseudo)gaps and at the same time can mediate Cooper pairing, splitting single bands in two branches, that in wave-vector space behave locally as valence (hole-like) and conduction (electron-like) bands. Finally, in FeSe it is possible to tune the position of the valence band with respect to the conduction band, and thus the carrier density, through the chemical potential alignment with the trilayer graphene substrate, where the local work function is spatially inhomogeneous \cite{Lin}.\\
In this work, we have investigated the superconducting state properties of a 2D electronic system with valence and a conduction bands considering the case of different intra-band and pair-exchange couplings, which is the typical configuration in bilayer graphene superconductors and electron-hole multilayer systems. We have obtained numerical results for the phase stiffness and the BKT critical temperature as functions of the energy band gap, the intra-band, and the pair-exchange couplings, that can be tuned in bilayer graphene systems by applying external electric fields through metal gates and, in the case of electron-hole multilayers, by tuning the insulating barrier width. We have found that there exist a giant enhancement of the Kosterlitz-Thouless critical temperature and a consequent shrinking of the pseudogap region, with a different mechanism from the one found in \cite{Salasnich2019}, in the regime of small energy gap between the bands and of low density of carriers in the conduction band, with the latter being the optimal condition for observing superfluidity in the electron-hole systems \cite{Perali2013, Conti2017}. 
Moreover, we have found a minimum in the  pseudogap region, in the intermediate-regime of the pair-exchange couplings, shifting to the weak-coupling regime by reducing the level of filling of the conduction band.
We also made comparison with single-band 2D superconducting systems, in which the phase stiffness is directly proportional to the electron density, resulting in small values of the BKT transition temperatures in the low density regime. It turns out that the presence of the valence band act as a reservoir of electrons for the two-band system, contributing constructively to the total stiffness of the system. Thus, we predict that in this regime, multi-band superconducting and superfluid systems with valence and conduction bands can have enhanced Kosterlitz-Thouless critical temperatures with respect to single-band and single-condensate systems.

The manuscript is organized as follow. In section \ref{sec1} we describe the model for the physical system considered and the theoretical approach for the evaluation of the superconducting state properties. In section \ref{sec2} we report our results. The conclusions of our work will be reported in Section \ref{sec3}.

\section{Model and Methods}
\label{sec1}

\begin{figure}[h]
\hspace{-0.2cm}\includegraphics[width=.48\textwidth]{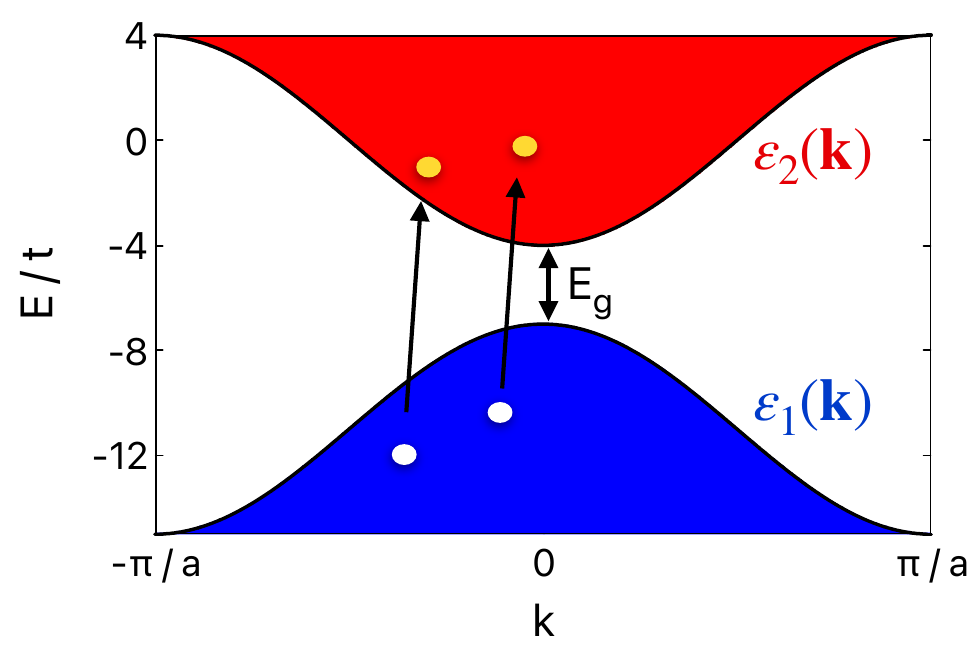}
\caption{Band dispersions for the two-band system. The energy and the wave- vectors are measured in units of $t$ and $a$, respectively. $E_g$ is the energy gap between the valence and the conduction bands.}
\label{fig1}
\end{figure}

We consider a two-dimensional (2D) two-band superconductor with a valence and a conduction electronic band in a square lattice. This model can be applied to 2D electron-hole layered superfluids as well. The valence and the conduction band have a tight-binding dispersion given, respectively, by Eqs. (\ref{eqn:11}) and (\ref{eqn:12}): 
\begin{equation}
\varepsilon_1(\mathbf{k})=2t[\cos(k_x a)+\cos(k_y a)]-8t-E_g
\label{eqn:11}
\end{equation}
\begin{equation}
\varepsilon_2(\mathbf{k})=-2t[\cos(k_x a)+\cos(k_y a)]
\label{eqn:12} 
\end{equation}
where $t$ is the nearest neighbour hopping parameter, $a$ the lattice constant, $E_g$ the energy band-gap between the conduction and valence band. and the wave-vectors belong to the first Brillouin zone $-\frac{\pi}{a} \leq k_{x,y} \leq \frac{\pi}{a}$. The band dispersions are reported in Fig. \ref{fig1}.
 We assume that Cooper pairs formation is due to an attractive interaction between opposite spin electrons. The two-particle interaction has been approximated by a separable potential $V_{ij}(\mathbf{k}, \mathbf{k'})$ with an energy cutoff $\omega_0$, which is given by:
\begin{equation}
V_{ij}(\mathbf{k}, \mathbf{k'})=-V_{ij}^{0} \Theta\Bigl(\omega_0-|\xi_i(\mathbf{k})|\Bigr) \Theta\Bigl(\omega_0-|\xi_j(\mathbf{k'})|\Bigr)
\label{eqn:13}
\end{equation}
where $V_{ij}^0>0$ is the strength of the potential and $i,j$ label the bands. $V_{11}^0$ and $V_{22}^0$ are the strength of the intra-band pairing interactions (Cooper pairs are created and destroyed in the same band).
$V_{12}^0$ and $V_{21}^0$ are the strength of the pair-exchange interactions (Cooper pairs are created in one band and destroyed in the other band, and vice versa), so that superconductivity in one band can induce superconductivity in the other band. We set $\hbar=1$ and $k_B=1$ throughout the manuscript.

The same energy cutoff $\omega_0$ of the interaction for intra-band and pair-exchange terms is considered, and is taken much larger than the bandwidth so that the interaction can be considered contact-like. The terms corresponding to Cooper pairs forming from electrons associated with different bands (inter-band or cross-band pairing) are not considered in this work (see \cite{Paredes2020}). The term $\xi_i(\mathbf{k})=\varepsilon_i(\mathbf{k})-\mu$ in Eq. (\ref{eqn:13}) is the energy dispersion with respect to the chemical potential $\mu$. The index $i = 1, 2$ numerates the bands, where $i = 1$ denotes the valence band and $i = 2$ the conduction band.

The superconducting state of the two-band system is examined within a mean field theory. We consider the two-band case generalization of the superconducting gap equation at finite temperature $T$
\begin{equation}
\Delta_i(\mathbf{k})=-\frac{1}{\Omega}\sum_{j} \sum_{\mathbf{k'}}V_{ij}(\mathbf{k}, \mathbf{k'})\frac{\tanh{\frac{E_j(\mathbf{k'})}{2 T}}}{2E_j(\mathbf{k'})} \Delta_j(\mathbf{k'})
\label{eqn:14}
\end{equation}
where $E_i(\mathbf{k})=\sqrt{{\xi_i(\mathbf{k})}^2+{\Delta_i(\mathbf{k})}^2}$ are excitation branches in the superconducting state, and $\Omega$ is the area occupied by the 2D system. Note that for a separable interaction of the form of Eq. (\ref{eqn:13}), the superconducting gaps assume the following expression:
\begin{equation}
\Delta_i(\mathbf{k})=\Delta_i\Theta\Bigl(\omega_0-|\xi_i(\mathbf{k})|\Bigr)
\label{eqn:16}
\end{equation}
We point out that all the coupling configurations considered in this work led to solutions of the gap equations in Eqs. \ref{eqn:14} which are global minima of the free energy and thus stable physical solutions for the two-gap superconducting state, as discussed in Ref.\cite{Aase2023}. 
The total electron density of the two-band system is fixed and it is given by the sum of the densities of the single bands, $n_{tot} =\sum_{i} n_i$, that can vary instead. The electronic density $n_i$ in the band ($i$) at a temperature T is defined as:
\begin{equation}
n_i=\frac{2}{\Omega}  \sum_{\mathbf{k}} \Bigl[{v_i(\mathbf{k})}^2 f\big(-E_i(\mathbf{k})\big)+{u_i(\mathbf{k})}^2 f\big(E_i(\mathbf{k})\big)\Bigr]
\label{eqn:17}
\end{equation}
where $f$ is the Fermi-Dirac distribution function. The BCS coherence weights $v_i(\mathbf{k})$ and $u_i(\mathbf{k})$ are:
\begin{equation}
{v_i(\mathbf{k})}^2=\frac{1}{2}\Bigg[1- \frac{\xi_i(\mathbf{k})}{\sqrt{{\xi_i(\mathbf{k})}^2+{\Delta_i(\mathbf{k})}^2}}\Bigg]
\label{eqn:18}
\end{equation}
\vspace{.2cm}
\begin{equation}
{u_i(\mathbf{k})}^2=1-{v_i(\mathbf{k})}^2
\label{eqn:19}
\end{equation}
The low-temperature physics of a 2D attractive Fermi gas turns out to be different from that of a 3D one. This is a consequence of the Mermin–Wagner theorem, which prohibits the spontaneous breaking of a continuous symmetry at finite temperatures, allowing one to find off-diagonal long-range order only at $T = 0$ in two dimensions. However, in the low temperature phase of these kind of 2D systems there exists a ”quasi–long range order” in which the phase correlations decays algebraically, i.e. $\langle e^{i (\theta(r)- \theta(0))}\rangle \sim |r|^{-\eta}$, where $\eta$ is a $T$ - dependent exponent and $\theta$ is the phase of the order parameter. This algebraic decay of the correlation function is observed up to a finite temperature $T_{KT}$, known as the Berezinskii-Kosterlitz-Thouless (BKT) critical temperature, that separates the low-temperature from the high-temperature phase. The high temperature phase is characterized instead by an exponential decay of phase correlation $\langle e^{i (\theta(r)- \theta(0))}\rangle \sim e^{-r/\xi}$, where $\xi$ is a characteristic length of the system and depends on temperature.

The transition temperature can be determined through the relation
\begin{equation}
T_{BKT}= \frac{\pi}{2} J_{TOT} (T_{BKT})
\label{eqn:20}
\end{equation}
where $J_{TOT}(T_{BKT})$ is the total phase stiffness at the Kosterlitz-Thouless transition temperature, that in our case is given by the sum of the stiffnesses of the single bands $J_{TOT}=\sum_{i} J_i$.\\
Following the approach used in Appendix A of Ref. \cite{Metzner2019}, the phase stiffness of a superconductor can be computed from the current density induced by an external electromagnetic field, given by
\begin{equation}
j_{i \alpha} (\mathbf{q}, \omega)=- \sum_{\alpha'} K_{i \alpha \alpha'}(\mathbf{q}, \omega) A_{\alpha'} (\mathbf{q}, \omega)
\label{eqn:21}
\end{equation}
where $A_{\alpha}$ are the component of the vector potential in a gauge where the scalar potential $\phi$ vanishes and $\nabla \cdot \mathbf{A} = 0$, $K_{i \alpha \alpha'}(\mathbf{q}, \omega)$ is the response function. The index $\alpha$ refers to the Cartesian axis and $i$ is the band index. The phase stiffness is related to the static limit of the response function,
\begin{equation}
 K_{i \alpha \alpha'}=\lim_{\mathbf{q} \to 0}  K_{i \alpha \alpha'}(\mathbf{q}, 0) 
\label{eqn:22}
\end{equation}
The response function in Eq. (\ref{eqn:21}) is given by the sum of a paramagnetic and a diamagnetic contribution,
\begin{equation}
K_{i \alpha \alpha'}=K_{i \alpha \alpha'}^{dia}+K_{i \alpha \alpha'}^{para}
\end{equation}
\begin{equation}
 K_{i \alpha \alpha'}^{para}= \frac{2e^2}{V} \sum_{\mathbf{k}} f'\big(E_i(\mathbf{k})\big) \varepsilon^{\alpha}_{i}(\mathbf{k}) \varepsilon^{ \alpha'}_{i}(\mathbf{k})
\label{eqn:23}
\end{equation}
\begin{equation}
 K_{i \alpha \alpha'}^{dia}= \frac{e^2}{V} \sum_{\mathbf{k}} \Biggl[ 1-\frac{\xi_i(\mathbf{k})}{E_i(\mathbf{k})}+\frac{2\xi_i(\mathbf{k})}{E_i(\mathbf{k})}f\big(E_i(\mathbf{k})\big) \Biggr] \varepsilon^{\alpha \alpha'}_{i}(\mathbf{k})
\label{eqn:24}
\end{equation}
with $\varepsilon^{ \alpha}_{i}(\mathbf{k}) = \partial{\varepsilon_i (\mathbf{k}})/\partial{k_{\alpha}}$, and $\varepsilon^{\alpha \alpha'}_{i}(\mathbf{k}) = \partial^2{\varepsilon_i (\mathbf{k})}/ \bigl(\partial{k_{\alpha}}\partial{k_{\alpha'}}\bigr)$
The off-diagonal elements ($\alpha \neq \alpha'$) of the response function vanish since the dispersions are symmetric in $k_x$ and $k_y$ because of the square symmetry of the considered lattice. Furthermore, the response function does not depends on the direction, that is, $K_{i \alpha}= K_i$ is independent of $\alpha$. The phase stiffness $J_i$ is independent on the direction as well, and is given by
\begin{equation}
 J_i=\frac{K_i}{(2e)^2 }
\label{eqn:23}
\end{equation}
The energies are normalized in units of the hopping parameter $t$ and the dimensionless couplings $\lambda_{ij}$ are defined as $\lambda_{ij}=N V_{ij}^0$, where $N=1/4 \pi a^2 t$ is the density of states at the top/ bottom of the valence / conduction band, that coincide since the density of states is not modified by the concavity of the band.

\section{Results}
\label{sec2}
In this section, we present results for the phase stiffness calculated at the BKT transition temperature and the BKT transition temperature itself, that are closely connected by the Kosterlitz-Nelson criterion in Eq.(\ref{eqn:20}), for the two-dimensional system made up by a valence and a conduction band. The superconducting order parameters $\Delta_1$ and $\Delta_2$, which enter the expressions for the total phase stiffness $J_{TOT}$ derived in Sec. \ref{sec1}, are computed from Eqs.(\ref{eqn:14}) coupled with Eq.(\ref{eqn:16}), that gives the total density of the system, considering a contact-type interaction ($\omega_0 / t =20$).

\begin{figure}[!b]
    \captionsetup[subfigure]{labelformat=empty}
    \centering
    \begin{subfigure}{0.4\textwidth}
        \includegraphics[width=\linewidth]{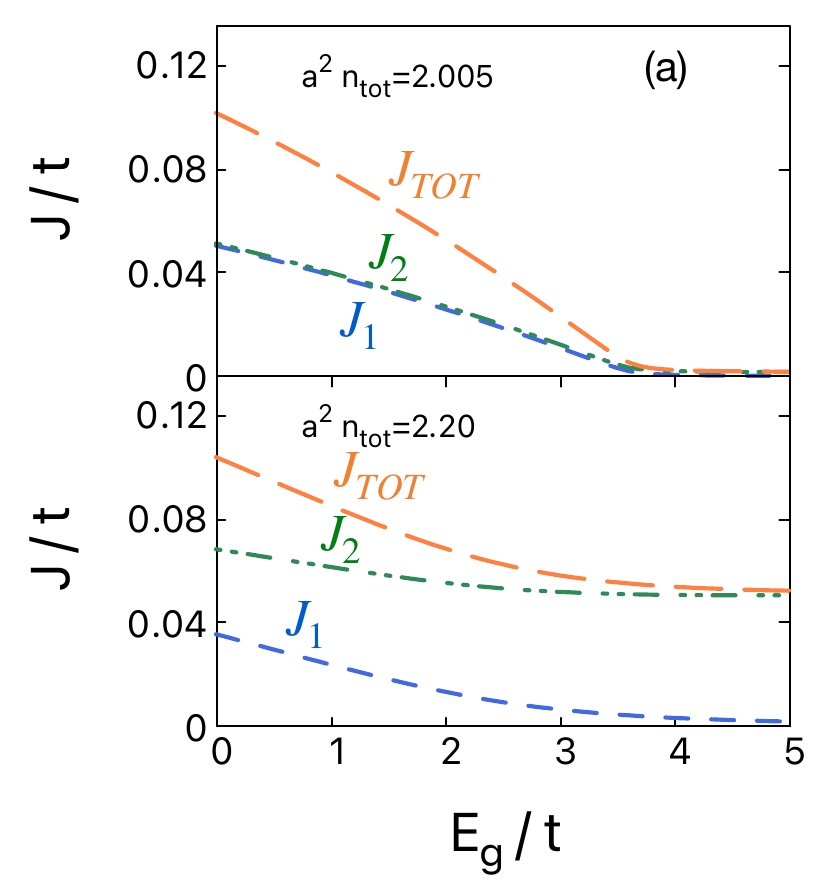}
        \subcaption{}
        \label{2a}
    \end{subfigure}
\medskip
    \begin{subfigure}{0.4\textwidth}
        \includegraphics[width=\linewidth]{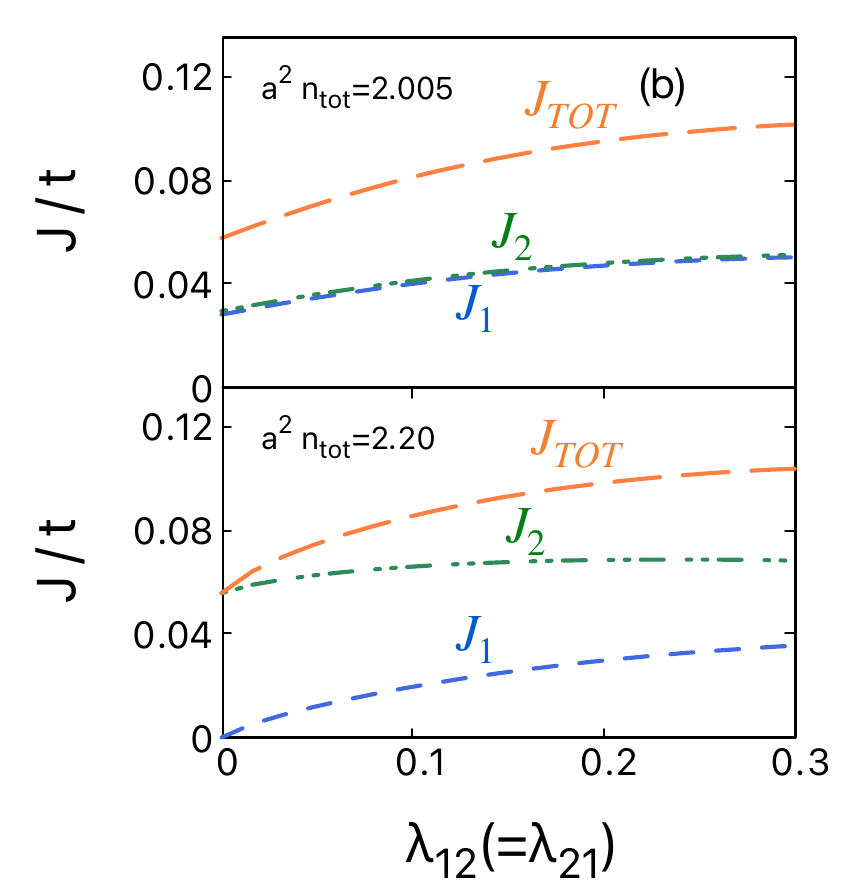}
        \subcaption{}
        \label{2b}
    \end{subfigure}
    \caption*{}
\end{figure}%
\begin{figure}[ht]\ContinuedFloat
    \captionsetup[subfigure]{labelformat=empty}
    \centering
    \begin{subfigure}{0.4\textwidth}
        \includegraphics[width=\linewidth]{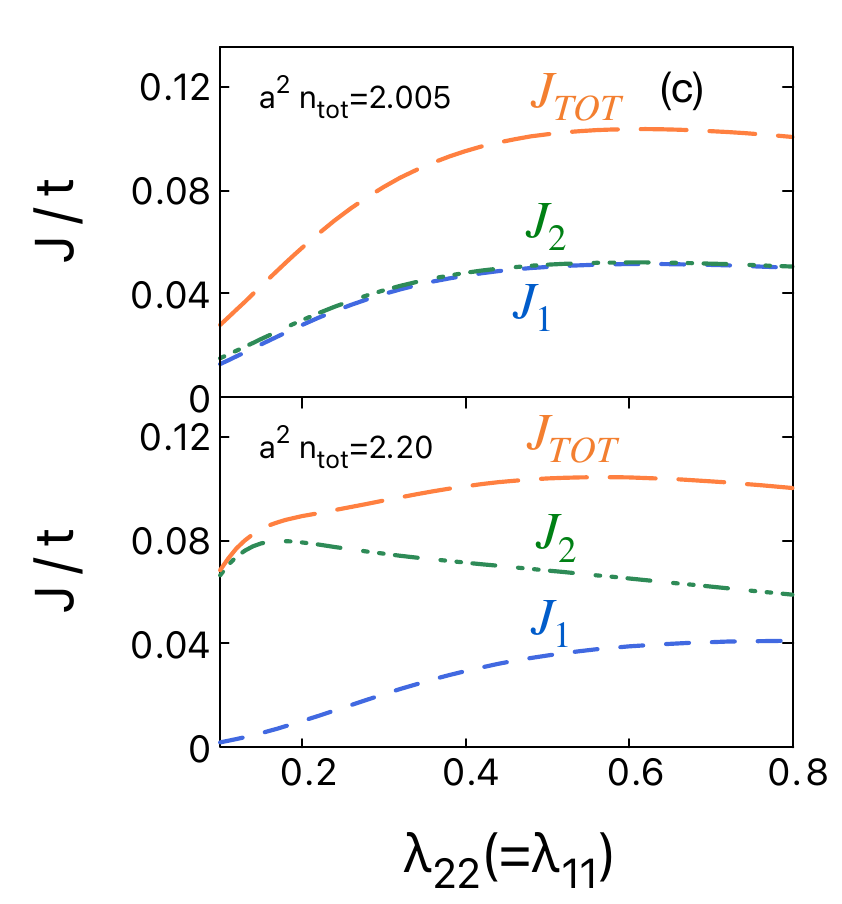}
        \subcaption{}
        \label{2c}
    \end{subfigure}
    \caption{Valence band $J_1$, conduction band $J_2$ and total phase stiffness $J_{TOT}=J_1+J_2$ as functions of $E_g / t$ (a), $\lambda_{12}=\lambda_{21}$ (b), and $\lambda_{22}=\lambda_{11}$ (c), calculated at $T=T_{BKT}$. The values of the total density are $a^2 n_{tot}=2.005$ (top panels) and $a^2 n_{tot}=2.20$ (lower panels). (a): $\lambda_{12}=\lambda_{21}=0.3$, $\lambda_{22}=\lambda_{11}=0.5$; (b): $E_g / t=0.0$; $\lambda_{22}=\lambda_{11}=0.5$; (c): $E_g / t=0.0$; $\lambda_{12}=\lambda_{21}=0.3$.}
    \label{fig2}
\end{figure}

We report the phase stiffness of single bands and the total phase stiffness as functions of the band-gap energy $E_g$ in Figs.\ref{2a}.
The presence of the valence band contributes to enhance the total stiffness of the system, especially in the region of small band-gap energy where the transfer of electrons from
the valence to the conduction band is favoured. 
The consequent presence of the holes in the valence band contributes constructively to 
\begin{figure}
 \captionsetup[subfigure]{labelformat=empty}
\centering
\begin{subfigure}{0.43\textwidth}
    \includegraphics[width=\textwidth]{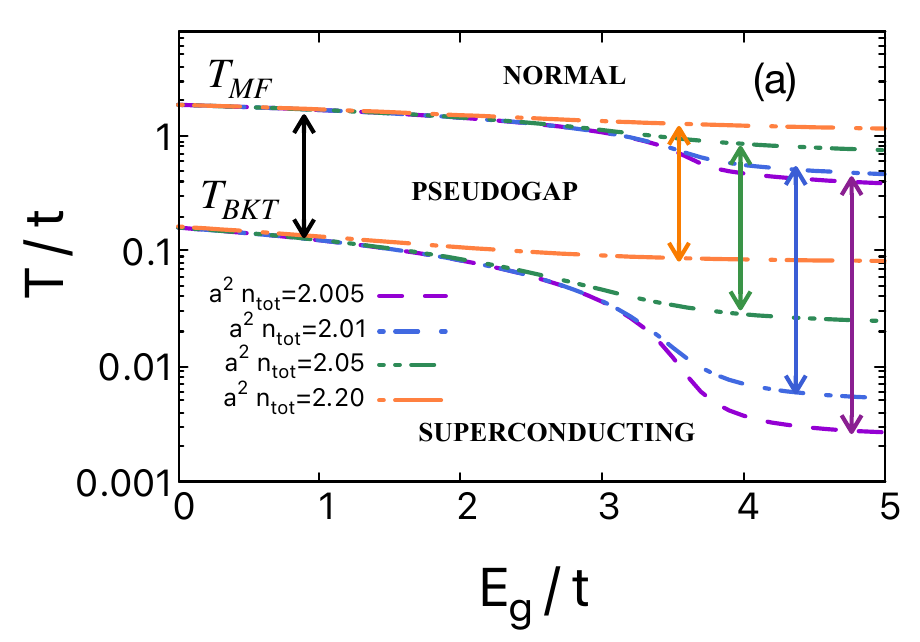}
    \subcaption{}
    \label{3a}
\end{subfigure}
\hfill
\begin{subfigure}{0.43\textwidth}
    \includegraphics[width=\textwidth]{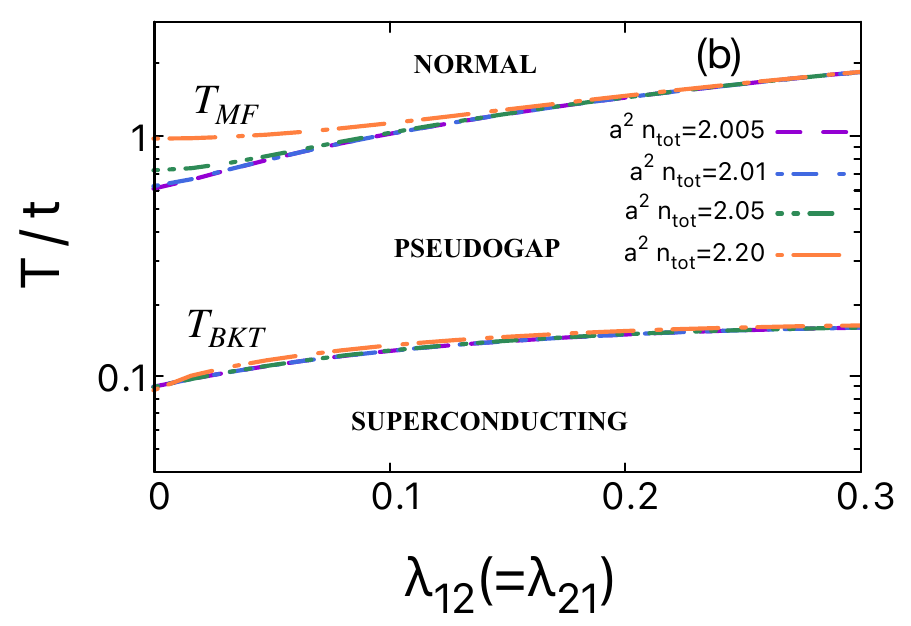}
     \subcaption{}
    \label{3b}
\end{subfigure}
\hfill
\begin{subfigure}{0.43\textwidth}
    \includegraphics[width=\textwidth]{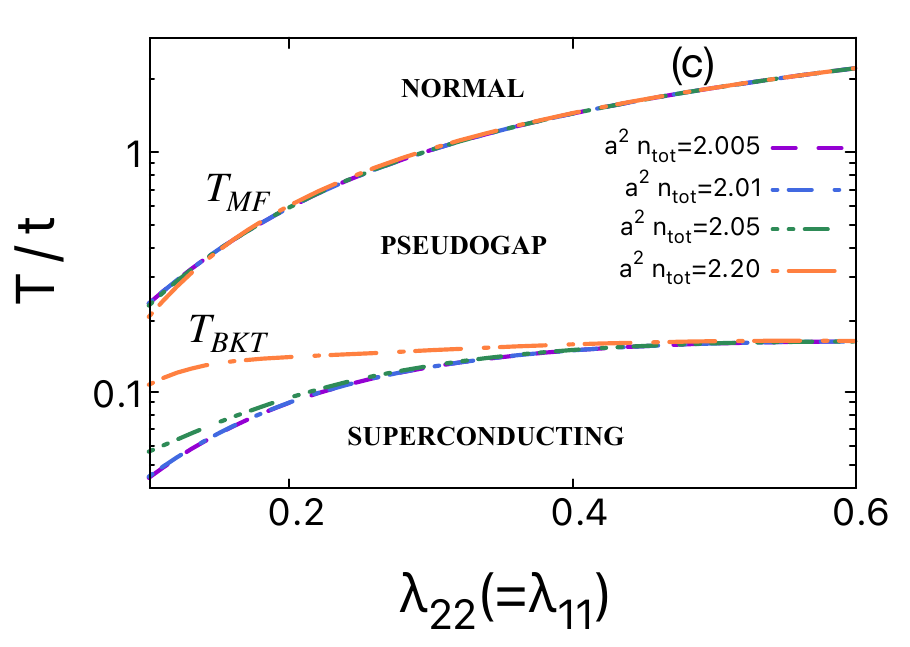}
   \subcaption{}
    \label{3c}
\end{subfigure}
\caption{Phase diagram of the system in the temperature $T / t$ vs band-gap energy $E_g / t$ (a), pair-exchange couplings $\lambda_{12}=\lambda_{21}$ (b) and intra-band couplings $\lambda_{22}=\lambda_{11}$ (c) planes, for different values of the total density.  (a): $\lambda_{12}=\lambda_{21}=0.3$, $\lambda_{22}=\lambda_{11}=0.5$; (b): $E_g / t=0.0$; $\lambda_{22}=\lambda_{11}=0.5$; (c): $E_g / t=0.0$; $\lambda_{12}=\lambda_{21}=0.3$.}
\label{fig3}
\end{figure}
the total stiffness of the system, adding up to the stiffness of the conduction band electrons and furthermore, there is a boost in the conduction band stiffness itself due to the electrons coming from the valence band. When the band-gap energy becomes too large the transfer of electrons is reduced, and superconductivity cannot exploit the presence of a proximate valence band, with the two condensates being essentially decoupled. 
 In this regime, when the density of the conduction band is small (top panel of Fig.\ref{2a}) superconductivity is strongly suppressed, since very few holes and electrons that can form Cooper-pairs are present in valence and conduction bands, respectively, thus leading to a suppression of the total phase stiffness $J_{TOT}$.
 When the density in the conduction band is higher (bottom panel of Fig.\ref{2a}), for large $E_g$ superconductivity is sustained only by the condensate in the conduction band, which is the only significant condensate in the system. This means that the enhancement in the total phase stiffness due to the valence band condensate is lost, since in absence of holes, the filled valence band will not create any stiffness and will be inactive for superconductivity. Thus, in this regime the system is acting as a single component condensate. In Figs.\ref{2b} we report the phase stiffness of the single bands and the total phase stiffness as functions of pair-exchange couplings $\lambda_{12}=\lambda_{21}$. For higher values of $\lambda_{12}=\lambda_{21}$ the total phase stiffness is enhanced, since the pair-exchange interactions facilitate the mixing of electrons between the two bands.
For small values of the conduction band electron density (top panel of Fig.\ref{2b}), even in the limit of zero pair-exchange interactions the presence of the valence band is enough to enhance the total stiffness with respect to a system having only the conduction band. In fact, 
\begin{figure}
 \captionsetup[subfigure]{labelformat=empty}
\centering
\begin{subfigure}{0.43\textwidth}
    \includegraphics[width=\textwidth]{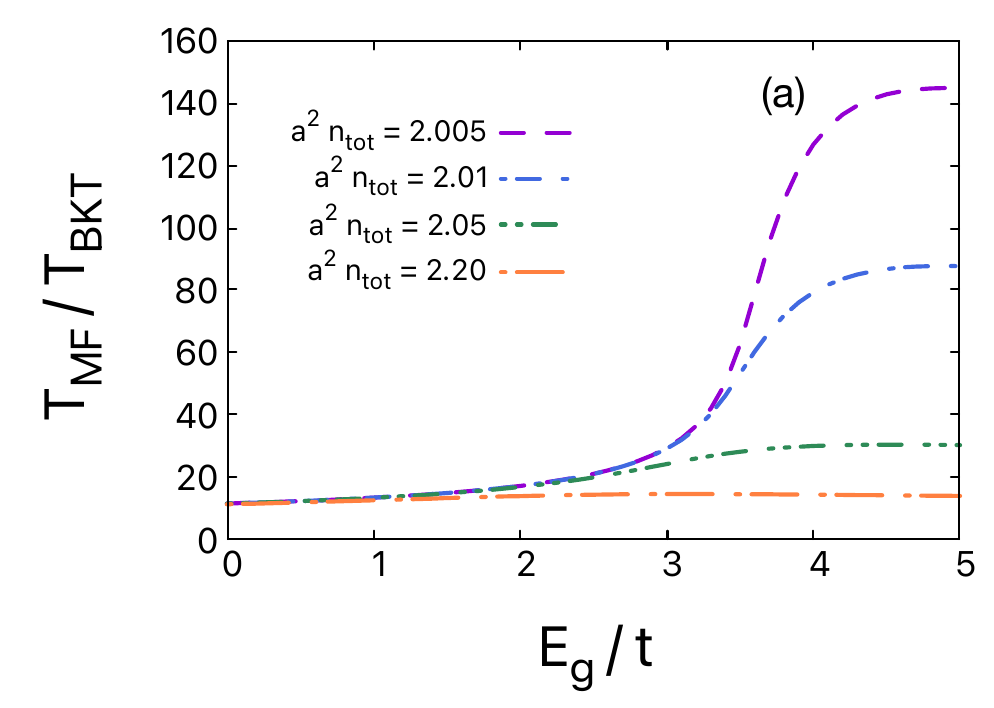}
  \subcaption{}   
    \label{4a}
\end{subfigure}
\hfill
\begin{subfigure}{0.43\textwidth}
    \includegraphics[width=\textwidth]{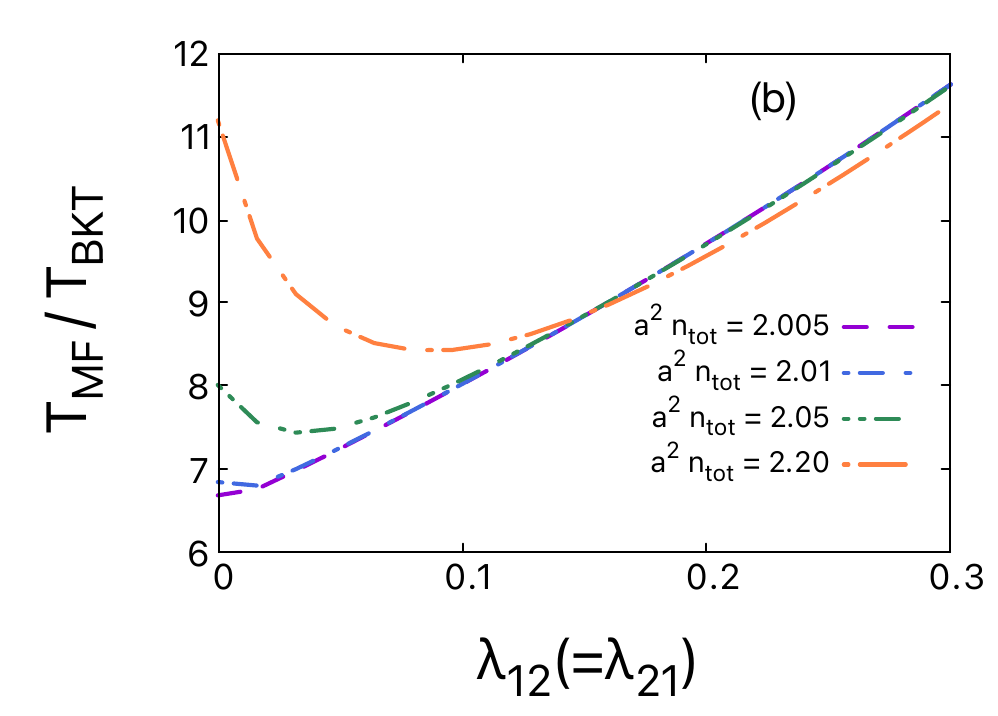}
   \subcaption{} 
    \label{4b}
\end{subfigure}
\hfill
\begin{subfigure}{0.43\textwidth}
    \includegraphics[width=\textwidth]{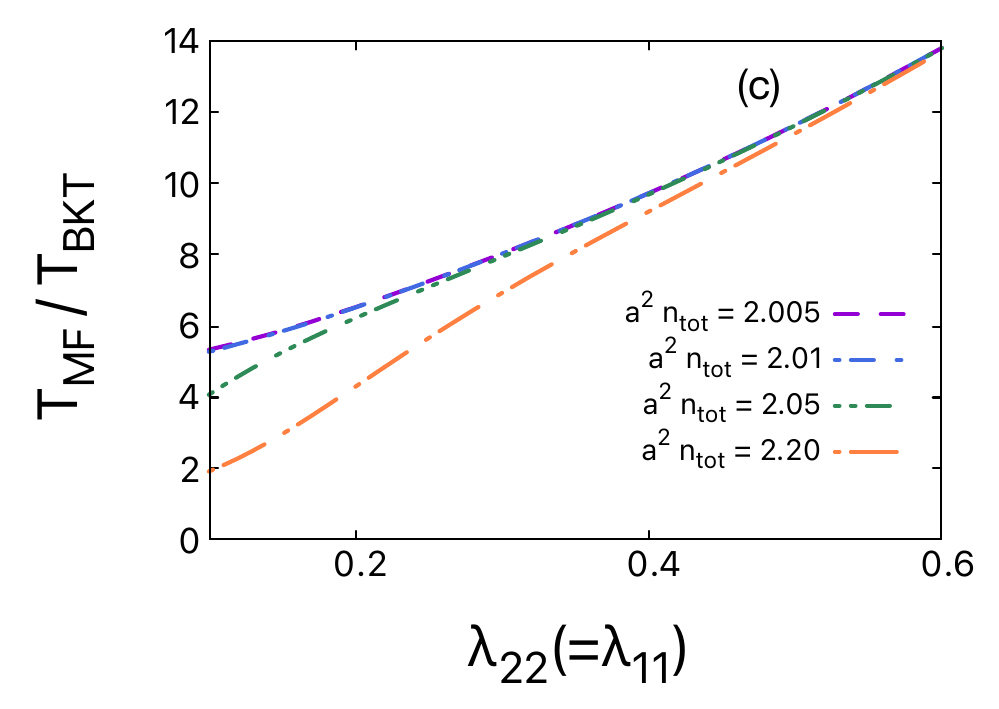}
   \subcaption{} 
    \label{4c}
\end{subfigure}
\caption{Ratio between the mean-field critical temperature $T_{MF}$ and the Berezinskii-Kosterlitz-Thouless critical temperature $T_{BKT}$ as a function of $E_g / t$ (a), $\lambda_{12}=\lambda_{21}$ (b), and $\lambda_{22}=\lambda_{11}$ (c), for different values of the total density.  (a): $\lambda_{12}=\lambda_{21}=0.3$, $\lambda_{22}=\lambda_{11}=0.5$; (b): $E_g / t=0.0$; $\lambda_{22}=\lambda_{11}=0.5$; (c): $E_g / t=0.0$; $\lambda_{12}=\lambda_{21}=0.3$.}
\label{fig4}
\end{figure}
the transfer of electrons from the valence band to the conduction band is not only due to the pair-exchange but to all the channels of the interaction, including also the intra-band channels. Moreover, since we are dealing with a system at finite temperature, thermal effects can also contribute to the redistribution of the elctrons between the two bands. However, for higher values of the conduction band electron density (lower panel of Fig. \ref{2b}), for zero pair-exchange couplings the intra-band channels alone are not enough to transfer the electrons in the conduction band, and the valence band condensate in not superconducting. Also in this regime the system is behaving as a single component condensate, with only the conduction band contributing to the total phase stiffness $J_{TOT}$. 
When the pair-exchange couplings are turned on, the Cooper pairs transfer from valence band to conduction band can start, leaving holes in the valence band, that in this way can host a superconducting condensate and give its contribution to the total stiffness. The physics is very similar to the previous case in the conduction band low density regime, when the intra-band couplings are tuned (top panel of Fig. \ref{2c}). For higher values of the conduction band electron density, (bottom panel of Fig. \ref{2c}), even though the stiffness of the conduction band condensate is again the only relevant contribution to the total stiffness in the weak-coupling regime, the stiffness of the valence band condensate is not completely zero as in the previous case, due to the presence of  finite pair-exchange interactions that couple the two condensates.
In the intermediate- and strong- coupling regime, the stiffness of the conduction band condensate is decreasing, but the total stiffness is sustained by the valence band condensate stiffness, that is increasing instead. 

The behavior of the phase stiffness described until now is reflected into the BKT transition temperature. In Fig. \ref{3a} the mean-field and the Kosterlitz-Thouless critical temperature are reported as functions of the band-gap energy $E_g$. Note that the mean-field critical temperature $T_{MF}$ has been evaluated from the linearized form of Eqs.(\ref{eqn:14}) in the limit of vanishing superconducting gaps. While the mean-field critical temperature remains finite for low filling of the conduction band when $E_g$ becomes large, the Kosterlitz-Thouless critical temperature is strongly suppressed since the total phase stiffness is going to zero in this regime. Conversely, when $E_g$ is small, there is a giant enhancement in the BKT critical temperature due to the valence and conduction band condensates additive contributions to the total system stiffness.
\begin{figure}
\captionsetup[subfigure]{labelformat=empty}
\centering
\begin{subfigure}{0.4\textwidth}
    \includegraphics[width=\textwidth]{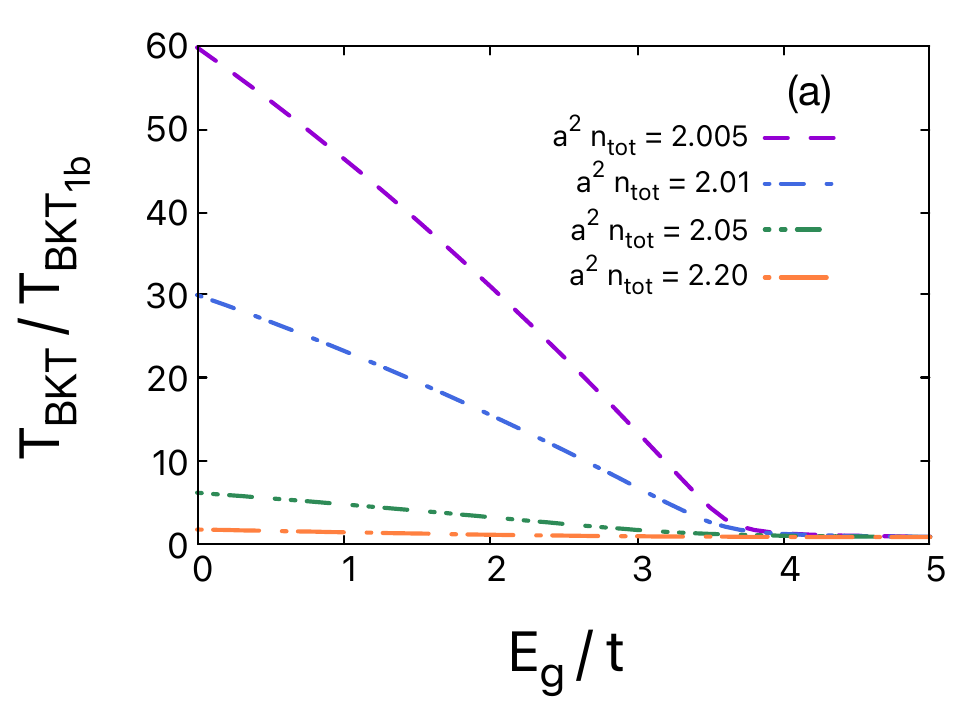}
  \subcaption{}   
    \label{5a}
\end{subfigure}
\hfill
\begin{subfigure}{0.4\textwidth}
    \includegraphics[width=\textwidth]{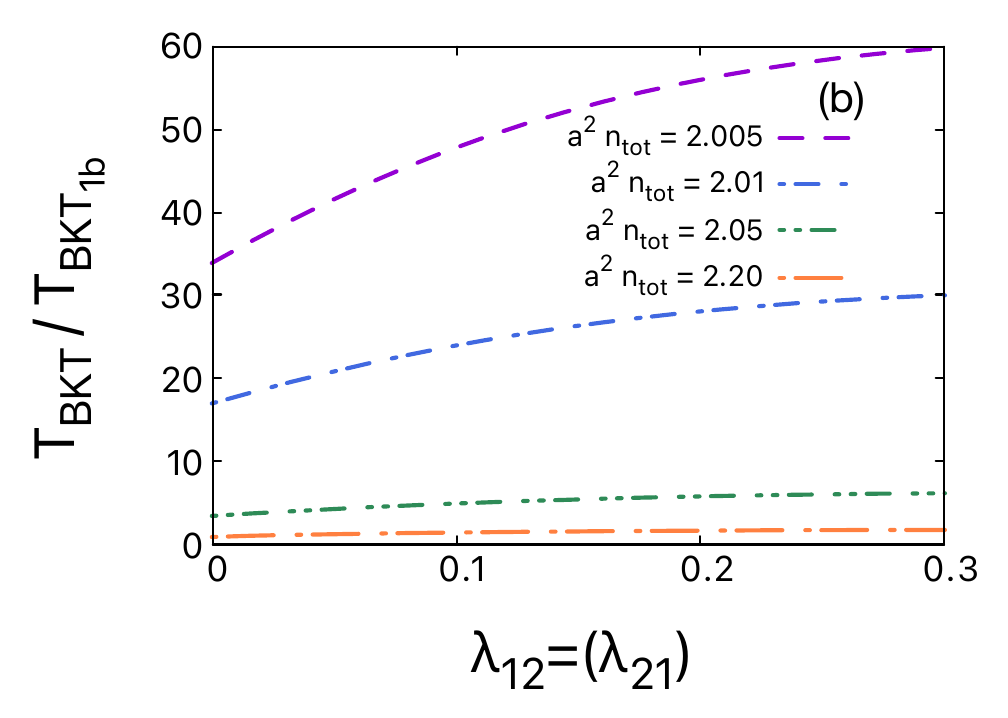}
  \subcaption{} 
    \label{5b}
\end{subfigure}
\hfill
\begin{subfigure}{0.43\textwidth}
    \includegraphics[width=\textwidth]{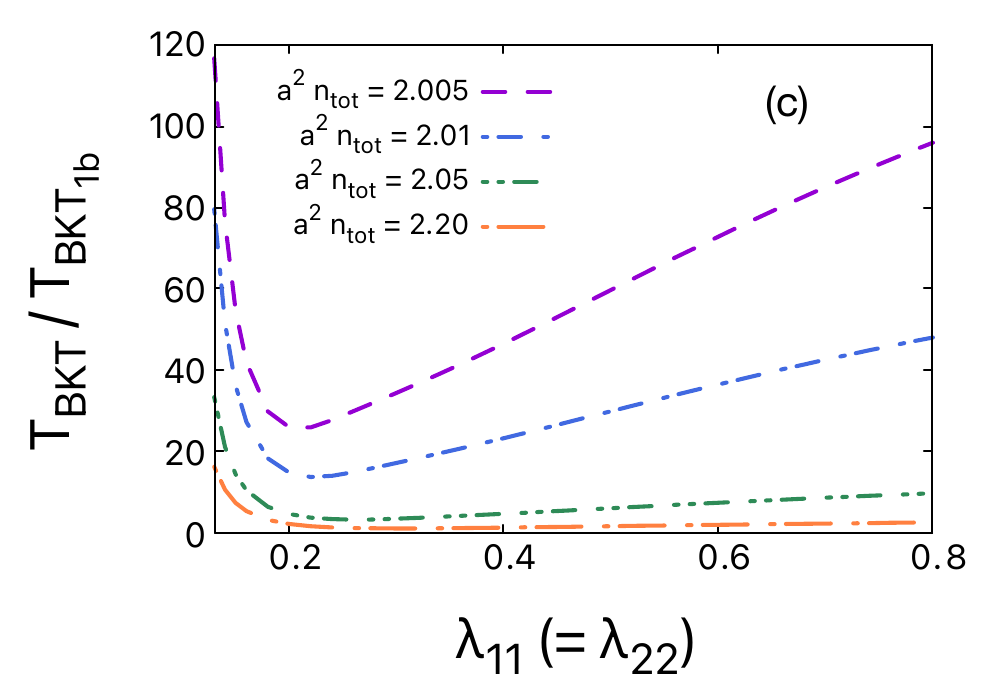}
    \subcaption{} 
    \label{5c}
\end{subfigure}
\caption{Ratio between the Berezinskii-Kosterlitz-Thouless critical temperature of the system with ($T_{BKT}$) and without ($T_{BKT_{1b}}$) valence band, as a function of $E_g / t$ (a), $\lambda_{12}=\lambda_{21}$ (b), and $\lambda_{22}=\lambda_{11}$ (c), for different values of the total density.  (a): $\lambda_{12}=\lambda_{21}=0.3$, $\lambda_{22}=\lambda_{11}=0.5$; (b): $E_g / t=0.0$; $\lambda_{22}=\lambda_{11}=0.5$; (c): $E_g / t=0.0$; $\lambda_{12}=\lambda_{21}=0.3$.}
\label{fig5}
\end{figure}
In Fig.\ref{3b} the mean-field and the Kosterlitz-Thouless critical temperature are reported as functions of the pair-exchange couplings $\lambda_{12}$. Both are enhanced for increasing $\lambda_{12}$, since the pair-exchange coupling induce the transfer of electrons from the valence band to the conduction band. However, while $T_{BKT}$ is almost insensitive to the level of filling of the conduction band, for small pair-exchange couplings values $T_{MF}$ is not. The situation is reversed when the intra-band couplings are tuned, as shown in Fig.\ref{3c}, with $T_{MF}$ that is insensitive to the level of filling of the conduction band while, in the weak-coupling regime $T_{BKT}$ is not, resulting in a suppression of superconductivity for decreasing values of the conduction band electron density.

Having a general overview on the phase diagram of the system, we now focus on the pseudogap region. In Fig. \ref{4a}, we show the ratio between the mean-field critical temperature $T_{MF}$ and the BKT transition temperature $T_{BKT}$ as a function of the energy band gap $E_g$. The ratio is almost constant in the region of small energy gap and for low density of carriers in the conduction band, while for larger $E_g$ the ratio is increased, evidencing an enhancement of the pseudo-gap region due to the decoupling of the two condensates. For higher values of carrier density, the behavior of the ratio is almost insensitive to the value of $E_g$ since the decoupling between the condensates occurs also for increasing number of electrons in the conduction band.
In Fig.\ref{4b}, we show the ratio between the mean-field critical temperature $T_{MF}$ and the BKT transition temperature $T_{BKT}$ as a function of
the pair-exchange couplings $\lambda_{21}=\lambda_{12}$. 
In this case for large values of the total density the ratio exhibits a non-monotonic behavior in the intermediate-coupling regime with the presence of a minimum in the pseudo-gap region. For larger values of the couplings the ratio keeps increasing as occurs in the conventional BCS theory, since $T_{BKT}$ saturates while $T_{MF}$ is going to infinity. For smaller values of the total density instead, the ratio is always monotonically increasing for all the values of the $\lambda_{21}=\lambda_{12}$ and the minimum in the pseudo-gap region is shifting toward the weak-coupling regime by further reducing the density. The ratio between $T_{MF}$ and $T_{BKT}$ as a function of the intra-band couplings in Fig.\ref{4c} instead, is always monotonically increasing for the different levels of filling of the conduction band, so that the minimum of the pseudogap region is always in the weak-coupling regime.

The next step is to study in detail the effect of the valence band on the BKT critical temperature. In Fig. \ref{5a} we show the ratio $T_{BKT} /T_{BKT_{1b}}$ between the BKT transition temperature of the two-band system and the BKT transition temperature of a system made up by a conduction band only, as a function of the band-gap energy. In the region of small $E_g$, the two condensate positively interfere and the BKT critical temperature is dramatically enhanced in the regime of low density, when the valence band is present with respect to the system made up by a conduction band only. 
The enhancement in the Kosterlitz-Thouless critical temperature can be observed also when the pair-exchange coupling is tuned, as shown in Fig.\ref{5b}. Surprisingly, we found that the amplification of $T_{BKT}$ with respect to the single band case is important even when the pair-exchange couplings are very small, for low levels of filling of the conduction band. This strong cooperative effect between the valence and conduction bands condensates avoids the suppression of the BKT transition temperature for low density of carriers, that occurs in the single band and single gap superconductors.
In Fig.\ref{5c}, we show the ratio $T_{BKT} /T_{BKT_{1b}}$ between the BKT transition temperature of our two-band system and the BKT transition temperature of the system made up by a conduction band only, as a function of the intra-band couplings. We have found again an amplification in the BKT critical temperature, which becomes huge in the regime of low filling of the conduction band. Moreover, in the weak-coupling regime the amplification presents a minimum, after which is monotonically increasing in the intermediate- to strong-coupling regime.

\section{Conclusions}
\label{sec3}
In this work, we have studied the BKT transition in a 2D superconducting electronic system with valence and conduction bands separated by a tunable energy gap. The electrons form Cooper pairs in the s-wave channel by interacting through an attractive potential with a large energy cutoff, which has been considered to model electronic interactions or electronic bands with relative small bandwidths. We have analyzed the behaviour of the phase stiffness and of the BKT critical temperature as functions of the energy band gap,  intra-band, and pair-exchange couplings for different levels of filling of the conduction band. We have found a giant enhancement of the BKT critical temperature in the regime of small energy gap between the bands and of small density of carriers in the conduction band. 

Alongside with the BKT transition temperature amplification, the pseudogap region between the mean-field temperature scale and the BKT transition temperature is suppressed in the conduction band low density regime by tuning the energy gap between the bands to small values, while is left unchanged for higher density values. By looking at the pair-exchange couplings instead, the pseudogap region has a non-monotonic behavior showing the presence of a minimum in the intermediate coupling regime, that shifts to the weak-coupling regime by reducing the filling of the valence band.

The pair-exchange and the intra-band couplings, favoured by small energy band gaps, induce the transfer of electrons from the valence band to the conduction band. The consequent presence of the holes in the valence band contributes constructively to the stiffness of the total system, adding up to the stiffness of the conduction band electrons that is boosted by the valence band electrons as well. In the absence of holes, the filled valence band will not create any stiffness and will be inactive for superconductivity. This strong cooperative effect avoids the suppression of the BKT transition temperature for low density of carriers, that occurs in single-band superconductors where only the conduction band is present. 

Despite the simplified nature of the pairing potential, our work gives a qualitative insight on the BKT transition in valence and conduction bands 2D superconducting systems and electron-hole superfluid systems, pointing toward optimal parameter ranges for amplification of the BKT transition temperature.

\section{Acknowledgments} 
This work has been supported by PNRR MUR project PE0000023-NQSTI.

\end{document}